\documentclass[11pt]{article}

\newsavebox{\PSLASH}
\sbox{\PSLASH}{$p$\hspace{-1.8mm}/}

\begin{document}
\title{Bessel Process and Conformal Quantum Mechanics }
\author{M. A. Rajabpour$^{a}$\footnote{e-mail: Rajabpour@to.infn.it} \\ \\
  $^{a}$Dip. di Fisica Teorica and INFN,
Universit{\`a} di Torino, Via P. Giuria 1, 10125 Torino,
\\Italy}
\maketitle
\begin{abstract}
Different aspects of the connection between the Bessel process and
the conformal quantum mechanics (CQM) are discussed. The meaning of
the possible generalizations of both models is investigated with
respect to the other model, including self adjoint extension of the
CQM. Some other generalizations such as the Bessel process in the
wide sense and radial Ornstein- Uhlenbeck process are discussed with
respect to the underlying conformal group structure.

\vspace{5mm}
\noindent PACS number(s):
\end{abstract}
\section{Introduction}

The Bessel process is one of the building blocks of stochastic
processes because of its applications and also its simplicity and
richness. The Bessel process describes the norm of the Brownian
motion in arbitrary dimension \cite{revuz yor,Borodin}. This process
describes the movement of an arbitrary point on the real line in the
Schramm-Loewner evolution \cite{schramm}. It is also important in
the probabilistic description of some financial markets specially
Cox, Ingersoll and Ross (CIR) model \cite{CoxIngersoll and Ross}.
Since the Bessel process is connected to the movement of a free
random walk in arbitrary dimension it could be connected to the
quantum mechanics of the free particle. The interpretation of
quantum mechanics as the classical stochastic equation is a long
story and was discussed in \cite{Edward Nelson}. It is based on
looking at the generator of the stochastic process as the
Hamiltonian of the quantum mechanics. Although by now this
interpretation is quite well known and was discussed also in many
books our case, the Bessel process, has not been discussed as much
as it deserves. The corresponding quantum mechanics is called
conformal quantum mechanics and it has many applications in
different areas of physics such as the Calogero model \cite{calegro}
, conformally invariant quantum mechanical models
 \cite{DFF,Jakiew2}, dynamics of quantum particles in the asymptotic
near-horizon region of black-holes \cite{CDKKT} and dipole-bound
anions as anisotropic conformal interaction \cite{camblong1}.
Although this quantum mechanics was discussed from many different
points of view, its connections and similarities with the Bessel
process have not been addressed so far and that is our goal in this
paper. From the stochastic processes point of view we have a
continuous process which is also scale invariant. The same
interpretation is true in the quantum mechanics part, but the story
in the quantum mechanics side can be generalized by considering some
special boundary conditions at the origin which can have stochastic
interpretations. It is also possible to define other generalized
Bessel processes which have interesting well-known quantum
mechanical counterparts. We will address different aspects of these
similarities and connections in this paper. The paper is organized
as follows:

In the second section we address many different properties of the
Bessel process, the starting point is our motivation from the norm
of the Brownian motion but we will soon generalize  the definition
in many different directions. In the third section by starting from
the quantum mechanics of the free particle we will extract conformal
quantum mechanics as a quantum mechanics with conformal symmetry.
Similar to the second section we will soon generalize the definition
by allowing also a real and even an imaginary number of dimensions.
Of course in every extension level we should be careful about the
physical meanings and possible interpretations with respect to the
original motivation. In this section we will also discuss the self
adjoint extension of the corresponding quantum mechanics. In the
fourth section we will discuss many connections and similarities
between Bessel process and CQM, in particular we explain the
possible meaning of the extension in each side with respect to the
other. This comparison will also propose the necessity of studying
generalized Bessel processes. In the fifth section we will
investigate the most natural and simple generalizations of Bessel
process including Bessel process with constant drift, Bessel process
in the wide sense, CIR model and Morse process which are related to
Coulomb potential, free particle conformal quantum mechanics, radial
harmonic oscillator and Morse potential respectively. We explain how
one can get the connections and also see the internal symmetries.
Finally in the last section we will summarize our results.

\section{Bessel Process}\
\setcounter{equation}{0}\ To define Bessel process we start with the
primary motivation for definition of the process. It is just the
radial part of $\delta$-dimensional Brownian motion, i.e,
$R_{t}=\sqrt{B^{2}_{1}(t)+...+B^{2}_{\delta}(t)}$. Using Ito's
formula one can write the following stochastic equation for the
Bessel process
\begin{eqnarray}\label{Bessel}
dR_{t}=\frac{\delta-1}{2R_{t}}dt+dB_{t}
\end{eqnarray}
where $B_{t}$ is the one dimensional Brownian motion. It is easy to
see that the above process has scaling property, i.e. if $R_{t}$ is
a Bessel process with starting point $x$ then the process
$c^{-1}R_{c^{2}t}$ is also Bessel process with starting point at
$x/c$ for positive $c$. In this level one can consider the above
equation with the arbitrary real $\delta$. The Bessel process with
positive $\delta$ was studied extensively in the literature
\cite{revuz yor,Borodin}. Before establishing the properties of
Bessel process it is worth mentioning some properties of Brownian
motion in arbitrary dimension.

It is well known that Brownian motion in $\delta\leq2$ is recurrent
and transient
  for $\delta>2$. With transient we mean that if $\delta>2$
then almost surly $\lim_{t\rightarrow\infty}|B_{t}|=+\infty$ and the
process is recurrent if the set $\{t:B_{t}\in U\}$ be unbounded for
all the sets $D$ in $R^{\delta}$; in other words the process $X_{t}$
said to be recurrent if $P_{x}(T_{y}<\infty)=1$ for all $x,y\in D$
where $T_{y}$ is the hitting time, the first time at which the
process hits point $y$, and $P_{x}(T_{y}<\infty)$ is the probability
of having finite hitting time of the point $y$ for the process with
starting point at $x$. Another interesting property of Brownian
motion is related to the probability of meeting two Brownian paths
with arbitrary starting points in finite time. Two Brownian paths
will meet each other if $\delta<4$ otherwise they will not
\cite{Lawler1}. In other words almost surely the probability of
intersection of two independent Brownian paths in $\delta\geq4$ is
zero.

Let's now summarize some of the properties of the Bessel process.
Feller's test, see appendix B, indicate that this process has a
natural boundary at infinity and a boundary at origin which is
natural if $\delta\geq2$, regular if $0<\delta<2$ and absorbing if
$\delta\leq0$. One can summarize also the following properties
\begin{eqnarray}\label{Bessel properties}
&I&: \mbox{for}\hspace{0.5cm} \delta>2 \hspace{1.5cm}\mbox{the Bessel process is transient},\nonumber\\
&II&: \mbox{for} \hspace{0.5cm}\delta\leq2 \hspace{1.5cm}\mbox{the Bessel process is recurrent},\nonumber\\
&III&: \mbox{for} \hspace{0.5cm}\delta\geq2\hspace{1.5cm} \mbox{the
origin is polar, it will not be touched
by the process},\nonumber\\
&IV&: \mbox{for}\hspace{0.5cm} 0<\delta<2,
\hspace{0.5cm}\mbox{origin could be a killing point or reflecting
point}\nonumber.
\end{eqnarray}

The same properties are true for the squared Bessel process defined
by $Z_{t}=R_{t}^{2}$ which satisfies the following equation
\begin{eqnarray}\label{SquaredBessel}
dZ_{t}=\delta (dt)+2\sqrt{|Z_{t}|}dB_{t}.
\end{eqnarray}
Using Ito's formula the generator of squared Bessel process is
\begin{eqnarray}\label{GeneratorSquareBessel}
\mathcal{L}f(x):=2xf''(x)+\delta f'(x).
\end{eqnarray}
Then one may introduce the Green's function $G_{\lambda}(x,y)$ as
the Laplace transform, with respect to time, of the transition
density, the density of finding the process starting from $x$ at
$y$, if of the process
\begin{eqnarray}\label{Laplace transform}
G_{\lambda}(x,y)=\int_{0}^{\infty}e^{-\lambda t}p(t,x,y).
\end{eqnarray}
The Green's function satisfies the following equation
\begin{eqnarray}\label{equation in Laplace space}
\mathcal{L}G_{\lambda}(x,y)-\lambda G_{\lambda}(x,y)=0.
\end{eqnarray}
The solution may be factorized as follows
\begin{eqnarray}\label{Green function}
G_{\lambda}(x,y)= \left\{
\begin{array}{l l}
  w^{-1}_{\lambda}\psi_{\lambda}(x)\phi_{\lambda}(y)  \quad \mbox{if $x\leq
y$}\\
  w^{-1}_{\lambda}\psi_{\lambda}(y)\phi_{\lambda}(x)  \quad \mbox{if $x\geq y$}\\
\end{array} \right .,
\end{eqnarray}
where $w_{\lambda}$ is the Wronskian
$w_{\lambda}:=\psi'_{\lambda}(x)\phi_{\lambda}(x)-\psi_{\lambda}(x)\phi'_{\lambda}(x)$.
Since in the case of $0<\delta<2$ the process is touching zero we
need to specify the boundary at the origin is a killing\footnote{In
some literatures the word absorbing was used} boundary condition or
reflecting boundary condition. Using the above boundary conditions
one could find the following solutions
\begin{eqnarray}\label{Solutions of Green function1}
\psi_{\lambda}(x)= \left\{
\begin{array}{l l}
  x^{-\frac{\nu}{2}}I_{\nu}(\sqrt{2\lambda
x})  \quad \mbox{if $\delta \geq 2$ or $0<\delta<2$ with reflecting origin}\\
  x^{-\frac{\nu}{2}}I_{-\nu}(\sqrt{2\lambda x})  \quad \mbox{if $\delta\leq0$ or $0<\delta<2$ with
killing origin,}\\
\end{array} \right .
\end{eqnarray}
and
\begin{eqnarray}\label{Solutions of Green function2}
\phi_{\lambda}(x)=x^{-\frac{\nu}{2}}K_{\nu}(\sqrt{2\lambda x}),
\end{eqnarray}
where $I_{\nu}$ and $K_{\nu}$ denote the modified Bessel functions
with index $\nu=\frac{\delta-2}{2}$. It is easy to see that
$w_{\lambda}=\frac{1}{2}$. Using the above solutions one can write
the transition density for all $\delta>0$ except the killing case as
follows
\begin{eqnarray}\label{transition density}
p(t,x,y)&=&\frac{1}{2t}(\frac{x}{y})^{-\frac{\nu}{2}}e^{-\frac{x+y}{2t}}I_{\nu}(\frac{\sqrt{xy}}{t})\hspace{0.3cm}\mbox{if}
\hspace{0.3cm}x>0,\\
p(t,x,y)&=&\frac{2}{(2t)^{1+\nu}\Gamma(1+\nu)}e^{-\frac{y}{2t}}\hspace{0.3cm}\mbox{if}\hspace{0.3cm}x=0.
\end{eqnarray}
The transition density for $0<\delta<2$ with the killing origin is
\begin{eqnarray}\label{transition density squared Bessel killing case}
p(t,x,y)=\frac{1}{t}(\frac{x}{y})^{-\frac{\nu}{2}}e^{-\frac{x+y}{2t}}I_{-\nu}(\frac{\sqrt{xy}}{t}).
\end{eqnarray}
Similar results could be calculated for the the Bessel process, in
this case the generator has the following form
\begin{eqnarray}\label{GeneratorSquareBessel}
\mathcal{L}f(x):=\frac{1}{2}f''(x)+\frac{1}{2x}(\delta-1) f'(x).
\end{eqnarray}
Since this process is just the square root of squared Bessel process
one can get the transition densities in this case by just the
transformations $x\rightarrow x^{2}$ and $y\rightarrow y^{2}$. For
example for the reflecting origin we will have
\begin{eqnarray}\label{transition density Bessel reflecting }
p(t,x,y)=\frac{1}{2t}y(\frac{x}{y})^{-\nu}e^{-\frac{x^{2}+y^{2}}{2t}}I_{\nu}(\frac{xy}{t}).
\end{eqnarray}
One could get the same answer by the method which is more familiar
for physicists and that is by using the Fokker-Planck equation which
has the following form for the Bessel process
\begin{eqnarray}\label{Fokker-Planck Bessel1}
\partial_{t}p(t,x,y)=\frac{1}{2}(\partial_{x}^{2}-\partial_{x}\frac{1+2\nu}{x})p(t,x,y).
\end{eqnarray}
One can do the transformation $p(t,x,y)=x^{\nu+\frac{1}{2}}Q(x,y,t)$
following by the Laplace transform and get the equation
\begin{eqnarray}\label{Fokker-Planck Bessel2}
-\frac{1}{2}\partial_{x}^{2}Q_{\lambda}+\frac{\nu^{2}-1/4}{2x^{2}}Q_{\lambda}=-\lambda
Q_{\lambda}.
\end{eqnarray}
where $Q_{\lambda}$ is the Laplace transform of $Q(x,y,t)$. The
above equation is just the modified Bessel equation with the
modified Bessel functions as solutions. It is easy to see that the
above eigenvalue problem by the change of variable
$S_{\lambda}=x^{-(\nu+1/2)}Q_{\lambda}$  is equivalent to
$(-\partial_{x}^{2}-(\frac{1+2\nu}{x})\partial_{x})S(x)=-\lambda
S(x)$ which the operator is the same as the generator of Bessel
process stated before.

So far we just addressed the Bessel process with natural boundary
conditions in origin and infinity but it is also possible to
investigate squared Bessel process with $\delta>0$ starting from the
$x\leq0$ or squared Bessel process with $\delta\leq0$ and arbitrary
starting point. These processes were studied in \cite{YorJaeschke}
and have the following properties: in the case $\delta=0$ the
process will reach zero and stays there. For a squared Bessel
process with $\delta\geq0$ and starting point $x\leq0$ one could
show that it behaves like $-Z_{-x}^{-\delta}$ process, with starting
point at $-x$, until it hits origin and after that it behaves like
$Z_{0}^{\delta}$. Similar relations are valid for the process with
$\delta\leq0$ and $x>0$ by just reversing the sign of dimension,
starting point and the process. For example the process with
negative dimension and negative starting point behaves as
$-Z_{-x}^{-\delta}$. It is worth to mention that since for negative
dimensions the squared Bessel process become negative the square
root of it, which is Bessel process, will become purely imaginary
and so one could be careful about extending the results to the
Bessel process, however up to the time that the process is positive
one could define the Bessel process as well as a real process. To
complete the discussion we give the transition density of the
squared Bessel process with negative dimension given in
\cite{YorJaeschke} for starting value $x>0$
\begin{eqnarray}\label{transition density for negative dimension}
p(t,x,y)&=&k(x,y,\delta,t)e^{-\frac{x+|y|}{2t}}\int_{0}^{\infty}\frac{(1+w)^{-\delta}}{w^{-\delta/2}}\exp{(\frac{-1}{2t}(x
w+\frac{|y|}{w}))};\nonumber\\
k(x,y,\delta,t)&=&\frac{-2^{\delta}}{\delta\Gamma^{2}[\frac{-\delta}{2}]}x^{1-\frac{\delta}{2}}|y|^{-(1+\frac{\delta}{2})}t^{\delta-1}.
\end{eqnarray}

Another interesting process related to the Bessel process is the
time reversed Bessel process. If $R_{t}$ be a Bessel process with
starting point on the positive real line with dimension smaller than
two then the time reversed Bessel process, defined by
$(R_{(T_{x\rightarrow0})-s},s\leq T_{x\rightarrow0})$, has the same
law with a Bessel process $\hat{R}_{s},s\leq \hat{L}_{0\rightarrow
x}$ starting from origin with $\hat{\delta}\equiv 4-\delta$ and
$\hat{L}_{0\rightarrow
x}\equiv\{\mbox{sup}\hspace{0.2cm}{t|\hat{R}_{t}}=x\}$. In the above
notation $T_{x\rightarrow y}\equiv \mbox{inf}\{t|R_{t}=y\}$ is the
first time that the process starting at $x$ hits the point $y$. To
make it more clear consider the ensemble of all Bessel paths that
ended up when they hit the origin in the first time, then consider
all these paths in the reverse time, the laws of these two processes
are the same. The other way to say the above statement is as
follows: For every bounded function $F$ one can write
\begin{eqnarray}\label{time reverse Bessel process}
E_{0}^{\delta}[F(R_{(T_{x\rightarrow0})-s},s\leq
T_{x\rightarrow0})]=E_{0}^{4-\delta}[F(\hat{R}_{s},s\leq
\hat{L}_{0\rightarrow x})],
\end{eqnarray}
where $F$ is a function on the set of realizations of the process.
The above equality relates Bessel process with dimension $\delta$ to
Bessel process with dimension $\delta'=4-\delta$. One can write this
duality by defining $\kappa=\frac{4}{1-\delta}$ as
$\frac{1}{\kappa}+\frac{1}{\kappa'}=\frac{1}{2}$. Using the equality
one could easily see that
$n=\sin(\frac{4\pi}{\kappa})=\sin(\frac{4\pi}{\kappa'})$. The above
equality is valid for all real values of $\delta$. It is also
possible to write the equality as $\nu+\hat{\nu}=0$, then one could
say that these two processes have the same $Q(x,y,t)$ if we work in
the positive range of dimensions of Bessel process. In fact we will
see in section two that these two different Bessel processes are
related to the same conformal quantum mechanics. It is also easy to
see that $\delta=2$ is self dual and $\delta=4$ is critical which
beyond that our reverse time process has negative dimension. The
above equality is also useful to get the law of certain first
hitting times of Bessel process \cite{YorJaeschke}.

\section{Conformal Quantum Mechanics}\
\setcounter{equation}{0}\

In order to introduce conformal quantum mechanics we follow the same
strategy that we had in the previous section, we start from the free
particle in $\delta$ dimensions. The Schr\"odinger equation for a
free particle in radial coordinates has the following form
\begin{eqnarray}\label{Schrodinger radial coordinate1}
-\frac{1}{2r^{\delta-1}}\frac{\partial}{\partial
r}(r^{\delta-1}\frac{\partial\varphi(r)}{\partial r})=E\varphi(r),
\end{eqnarray}
or after differentiation one could write the following Bessel
generator eigenvalue equation
\begin{eqnarray}\label{Schrodinger radial coordinate2}
-\frac{\partial^{2}\varphi(r)}{\partial
r^{2}}-\frac{\delta-1}{r}\frac{\partial\varphi(r)}{\partial
r}=2E\varphi(r).
\end{eqnarray}
As it is evident a similar equation was derived for the Bessel
process in Laplace space in the previous section. By
$\varphi(r)=r^{-(\nu+1/2)}Q(r)$ and extending the range of $\delta$
to all real values we will have the following Hamiltonian for
conformal quantum mechanics
\begin{eqnarray}\label{Conformal Quantum Mechanics Hamiltonian}
H=\frac{1}{2}(p^{2}+\frac{\nu^{2}-1/4}{r^{2}}),
\end{eqnarray}
where $p=i\frac{\partial}{\partial r}$ and $[r,p]=i$ at the quantum
level. The above quantum mechanics as a singular quantum mechanics
was discussed extensively soon after the discovery of quantum
mechanics \cite{Case,Frank} and references therein. It is easy also
to get the same quantum mechanics for two free particles moving in
$\delta$ dimensions with $\frac{1}{r^{2}}$ interaction by just
forgetting the momentum of the center of mass term and going to the
radial part of the spherical coordinates. The conformal quantum
mechanics has many interesting properties. Before investigating
different solutions of the above system we study some symmetries of
the system. The Hamiltonian at the level of one dimensional quantum
mechanics with usual inner product for $Q(r)$ is symmetric with
respect to $\nu\rightarrow-\nu$ but at the level of $\varphi(r)$ is
not symmetric with respect to the same transformation. The action
corresponding to the Hamiltonian (\ref{Conformal Quantum Mechanics
Hamiltonian}) is invariant under the following transformations
\begin{eqnarray}\label{Conformal Transformationsn}
t'&=&\frac{a t+b}{c t+d},\nonumber\\r'(t')&=&\frac{r(t)}{c t+d}
\hspace{1cm} \mbox{with}\hspace{1cm}ad-bc=1.
\end{eqnarray}
The above transformations are the conformal transformations in $0+1$
dimensions \cite{DFF}. The basic transformations are time
translation, dilation and special conformal transformation with the
following Noether charges
\begin{eqnarray}\label{Conformal Transformationsn and Generators}
t'&=&t+b,\hspace{1cm}H=H=\frac{1}{2}(p^{2}+\frac{\nu^{2}-1/4}{r^{2}});\\
t'&=&\alpha^{2}t,\hspace{1.2cm}D=tH-\frac{1}{4}(rp+pr);\\
t'&=&\frac{t}{ct+1}\hspace{1cm}K=t^{2}H-\frac{1}{2}t(rp+pr)+\frac{1}{2}r^{2}.
\end{eqnarray}
The above generators verify the algebra of the conformal group
$SO(1,2)$ which is
\begin{eqnarray}\label{Conformal Algebra1}
[H,D]=iH,\hspace{1cm} [D,K]=i K,\hspace{1cm} [H,K]=2i D.
\end{eqnarray}
Using the new definitions $L_{0}=\frac{1}{2}(H+K)$ and
$L_{\pm1}=\frac{1}{2}(H-K\pm2iD)$ one can write the algebra in the
more familiar form
\begin{eqnarray}\label{Conformal Algebra2}
[L_{0},L_{\pm1}]=\mp L_{\pm1},\hspace{2cm} [L_{+1},L_{-1}]=2L_{0}.
\end{eqnarray}
Explicit dependence of $K$ and $D$ on $t$ will guarantee their
conservations. The important thing to mention is that the argument
is not considering the most general case because we already knew
that the action is invariant up to a total derivative which could be
non zero in the presence of the boundary in the origin. We will
discuss both conformal invariant and anomalous case. This will be
more clear when we discuss the self adjoint extension of the quantum
mechanics. Another important thing to mention is although it seems
that we found three conservation laws for our two-dimensional phase
system, it is not difficult to check that they are in fact related
by the following relation
\begin{eqnarray}\label{Constraint}
HK+KH-2D^{2}=\frac{\nu^{2}-\frac{1}{4}}{2}-\frac{3}{8},
\end{eqnarray}
which is also the Casimir operator of the group. In order to present
different aspects of the above quantum mechanics we introduce
another variable
\begin{eqnarray}\label{g}
g=\nu^{2}-\frac{1}{4}.
\end{eqnarray}
Our quantum mechanics has different properties with respect to the
value of $g$, some of which we will summarize in the following. For
an arbitrary value of $g$ it is easy to show that if one could find
one bound state with energy $E$ then scaling the position by an
arbitrary factor $\alpha$ it is easy to construct a new solution
with energy $\alpha^{2}E$, which means that if there exist any bound
states then there is a bound state for every negative energy. This
is a direct consequence of the scaling property of this model. The
same story is true for the positive energy solutions of the model
which means that we have just planar waves with all the possible
positive energies which was also used to obtain equation
(\ref{Fokker-Planck Bessel2}). In the previous section we considered
$\lim_{r\rightarrow0} \varphi(r)=0$ for the killing boundary
condition and for the reflecting case we had $\lim_{r\rightarrow0}
\frac{\partial\varphi(r)}{\partial r}=0$. In the level of
conformally invariant quantum mechanics we will ignore the negative
energy solutions which means that our Hilbert space is made by wave
functions corresponding to continuous positive energy solutions. The
corresponding boundary condition also will be discussed in the end
of this section.

The Hamiltonian could have other solutions dependent on the value of
the coupling $g$. Firstly we should mention that the Laplacian
operator that was the starting point is self adjoint if we consider
it in the whole space. But one could extract other solutions by
removing the origin from the domain and considering the self adjoint
extension of the operator in the new domain. Of course existence of
the extension is dependent on the value of $g$ or in other words to
the corresponding dimension. All the necessary aspects of theory of
self adjoint extension were discussed in the appendix A. In the case
of our quantum mechanics the possible extensions were discussed in
\cite{pym,Gupta,Griffits}.

For $g\geq\frac{3}{4}$ the Hamiltonian is self adjoint with removed
origin now and has a scattering sector with the following solutions
\begin{eqnarray}\label{Bessel function solutions1}
\varphi(r)=(\sqrt{2E}r)^{\frac{1}{2}}J_{\nu}(\sqrt{2E}r)\hspace{0.5cm}or\hspace{0.5cm}(\sqrt{2E}r)^{\frac{1}{2}}Y_{\nu}(\sqrt{2E}r),
\end{eqnarray}
where $J$ and $Y$ are the Bessel functions. For $\nu\geq1$ just $J$
can be considered because in this case $Y$ at the origin is infinity
and for $\nu\leq-1$ just $Y$ has desired property. For $\nu\geq1$
which is equivalent to $\delta\geq4$ one could argue that it is not
possible to have a bound state solution because always we can just
consider one of the solutions of the Bessel equation. The boundary
condition in this case is $\lim_{r\rightarrow0} \varphi(r)=0$ and
can not be extended.

For $g<\frac{3}{4}$ since the Hamiltonian is not self adjoint one
can find a required self adjoint extension. It is better to
distinguish between the domain $\frac{-1}{4}\leq g<\frac{3}{4}$ and
$g<\frac{-1}{4}$. First we discuss the domain $\frac{-1}{4}\leq
g<\frac{3}{4}$ which is equivalent to $-1<\nu<1$ or $0<\delta<4$. In
this range the Hamiltonian requires a self adjoint extension with
the self adjoint parameter $z$ which is responsible to map two
deficiency subspaces by the unitary map $e^{iz}$. The important
point to mention is that although the Hamiltonian in the domain
$D(H)\equiv\{\varphi(0)=\varphi'(0)=0\}$ is not self adjoint, it is
still Hermitian.

If we consider $H^{\dag}$ as the adjoint of $H$, with the same
differential representation as $H$, then from the Von Neumann's
theory of deficiency indices we know that the deficiency subspaces
$K_{\pm}$ are made by the square integrable solutions of the
equation $H^{*}\phi_{\pm}=\pm i\phi_{\pm}$ in the desired domain. In
our case the solutions are
\begin{eqnarray}\label{Self adjoint solutions}
\phi_{+}(r)&=&r^{\frac{1}{2}}H_{\nu}^{1}(re^{i\frac{\pi}{4}}),\\\phi_{-}(r)&=&r^{\frac{1}{2}}H_{\nu}^{1}(re^{-i\frac{\pi}{4}}),
\end{eqnarray}
where $H_{\nu}^{1}$ is the Henkel function and both of
$\phi_{\pm}(r)$ are square integrable in the half line. Using the
above solutions one could argue that the Hamiltonian is self adjoint
in the domain
$D_{z}(H)=D(H)\oplus\{u(\phi_{+}(r)+e^{iz}\phi_{-}(r))\}$ where $u$
is an arbitrary complex number. To find the valid boundary condition
we need $\psi$ to be in the self adjoint domain. Consider
$\Phi=\phi_{+}(r)+e^{iz}\phi_{-}(r)$ then $\psi$ is in the self
adjoint domain if $<\Phi|H\psi>=<H\Phi|\psi>$ or
\begin{eqnarray}\label{Self adjoint relation}
\lim_{r\rightarrow0}[\Phi^{*}\frac{d\psi}{dr}-\psi\frac{d\Phi^{*}}{dr}]=0.
\end{eqnarray}
To check this we need the behavior of $\Phi$ close to the origin
\begin{eqnarray}\label{Self adjoint relation}
\Phi(r)&\rightarrow&
\frac{i}{\sin(\pi\nu)}[Ar^{\nu+\frac{1}{2}}+Br^{-\nu+\frac{1}{2}}],
\end{eqnarray}
where
\begin{eqnarray}\label{AB}
A&=&\frac{e^{-i\frac{3\pi\nu}{4}}-e^{i(z+\frac{3\pi\nu}{4})}}{2^{\nu}\Gamma(1+\nu)},\nonumber\\
B&=&\frac{e^{i(z+\frac{\pi\nu}{4})}-e^{-i\frac{\pi\nu}{4}}}{2^{-\nu}\Gamma(1-\nu)},
\end{eqnarray}
for $\nu\neq0$. Using the above relations one can write the equation
(\ref{Self adjoint relation}) for the boundary condition as follows
\begin{eqnarray}\label{Self adjoint boundary condition1}
(Ar^{\nu+1/2}+Br^{-\nu+1/2})\frac{d\psi}{dr}-(A(\nu+\frac{1}{2})r^{\nu-1/2}+B(\frac{1}{2}-\nu)r^{-\nu-1/2})\psi\rightarrow0.
\end{eqnarray}
The equation for $\nu=\frac{1}{2}$ is like
$B\frac{d\psi}{dr}|_{0}=A\psi(0)$. Actually $\nu=\frac{1}{2}$ is
very interesting because firstly it corresponds to $\delta=3$ and
also because it describes the possible boundary conditions for the
free particle in the half line. The simplicity of the results helps
us to investigate the possible meanings of the above self adjoint
extension with respect to the Brownian motion and the Bessel process
which we will discuss in more detail in the next section. The next
important case which deserves separate calculations is $\nu=0$. In
this case the function $\Phi$ has different properties near the
origin. Using  $\phi_{+}(r)=\phi^{*}_{-}(r)$ and
\begin{eqnarray}\label{nio=0 case}
\Phi(r)\rightarrow
\frac{2i}{\pi}r^{\frac{1}{2}}\ln(r)+[\frac{1}{2}+\frac{2i}{\pi}(\gamma-\ln2)]r^{\frac{1}{2}}+e^{iz}cc\nonumber\\
=(A+A^{*}e^{iz})r^{\frac{1}{2}}\ln(r)+(B+B^{*}e^{iz})r^{\frac{1}{2}}.
\end{eqnarray}
where $\gamma$ is the Euler constant and $cc$ is the conjugate of
the first term. One can find the following boundary condition
\begin{eqnarray}\label{nio=0 case boundary condition}
\mathbf{A}r^{\frac{1}{2}}\ln(r)\frac{d\psi}{dr}-\frac{1}{\sqrt{r}}[\mathbf{A}(1+\frac{\ln
r}{2}+1)+\frac{\mathbf{B}}{2}]\psi\rightarrow0,
\end{eqnarray}
where
\begin{eqnarray}\label{ABs}
\mathbf{A}=A+A^{*}e^{iz}\hspace{1cm}\mathbf{B}=B+B^{*}e^{iz}.
\end{eqnarray}
One can extend the results also for the case with $g<-\frac{1}{4}$
which is related to pure complex $\nu$. In this case we have still a
one parameter self adjoint extension and all of the results are
still valid if we put imaginary $\nu$ in to the formulas, in
particular in to the (\ref{Self adjoint boundary condition1}). The
important point to mention is in fact that by introducing the self
adjoint parameter we are classifying all the possible physical
boundary conditions of our system. In our case we should mention
that after the self adjoint extension our Hamiltonian is not
necessarily scale invariant and so could have a negative ground
state which is in fact the case here. Of course the ground state
depends on the self adjoint extension parameter. The energy and wave
function for our Hamiltonian were discussed before in \cite{Gupta}
and references therein, and the results are as follows.

The Hamiltonian does not have any bound state for $g\geq\frac{3}{4}$
but admits one bound state in the range $\frac{-1}{4}\leq g
<\frac{3}{4}$.  The wave function of the bound state up to the
normalization constant has the following form
\begin{eqnarray}\label{bound states wave function}
\psi(r)\sim r^{\frac{1}{2}}K_{\nu}(\sqrt{-2E_{\nu}}r),
\end{eqnarray}
with energies
\begin{eqnarray}\label{bound states energies}
E_{\nu\neq0}&=&-\bigl(\frac{\sin(\frac{z}{2}+\frac{3\pi
\nu}{4})}{\sin(\frac{z}{2}+\frac{\pi
\nu}{4})}\bigr)^{\frac{1}{\nu}},\\E_{\nu=0}&=&-\exp{\frac{\pi}{2}\cot(\frac{z}{2})}.
\end{eqnarray}

 For the case $g<\frac{-1}{4}$ we have infinite
discrete bound states \cite{Case} with the same wavefunction as
above but with imaginary $\nu$. To get the energies we just need to
put the wave function in the corresponding boundary condition which
yields
\begin{eqnarray}\label{infinite bound states}
E=-2(\frac{A}{B})^{\nu}\exp\{2\arg\Gamma(1+\nu)-2n\pi\},\hspace{0.5cm}\mbox{for}\hspace{1cm}
n\in \mathbf{Z}.
\end{eqnarray}
It is worth to emphasize again that since after the self adjoint
extension the action is not scale invariant we do not need to worry
about having a bound state because it does not enforce any other
bound state by the scaling argument.

One could summarize this section as follows: we introduced conformal
quantum mechanics as the quantum mechanics of the free particle in
$\delta$ dimensions. Constraining the domain of the quantum
mechanics to the space without origin forces new boundary conditions
to the system. By these boundary conditions for some values of $g$
we need to find a self adjoint extension of the Hamiltonian. The
method of extension helps to find all the relevant boundary
conditions and so the physics of the model is related drastically to
the self adjoint extension parameter. The extended Hamiltonian which
is not scale invariant admits discrete bound states. Since after the
self adjoint extension our conformal symmetry has some anomalies,
may be the expression conformal quantum mechanics is a bit confusing
but because of our first motivation we will keep using it. Different
aspects of this anomaly were discussed extensively in
\cite{kumar,camblong1,camblong2,camblong3} from the re-normalization
group point of view. The above results could be also explained from
the framework of two free particles moving in $\delta$ dimensions.
The self adjoint extension is just giving boundary conditions
corresponding to the time that two particles meet each other. For
example, bound states of the above system could be seen as the bound
states of two particles which behave like one particle. In the next
section we give some hints on how one can translate the above
results in the Brownian motion language.

\section{Bessel Process and CQM}\
\setcounter{equation}{0}\

In this section we are interested in discussing different features
of the similarity between Brownian motion in $\delta$ dimension and
free particles from a quantum mechanics point of view and also
similarities between Bessel process and CQM.

The first remark is for the Bessel process with positive dimension.
We had origin and infinity as the natural boundary conditions which
is also the case for our CQM. In addition before extending the
Hamiltonian we have a scale invariant system which is the case also
in the Bessel process. Just as we remarked in the previous section,
keeping the scaling symmetry of the CQM gives us an unbounded
continuous spectrum which is also true if we try to find the
transition density of the Bessel process. To make more clear the
connection between two models one can use the following relation
between the transition density of the Bessel process and the path
integral of CQM
\begin{eqnarray}\label{Pand S}
p(x,y,t)&=&\int_{q(t_{0})=x}^{q(t)=y}[dq(\tau)]\exp[-S(q(\tau))]\\
S(q)&=&\int\frac{1}{2}(\dot{q}+\frac{1-\delta}{2q})^{2}dt
\end{eqnarray}
It is not difficult to see that by a transformation
$p(t,x,y)=x^{\nu+\frac{1}{2}}Q(x,y,t)$ one can go from the above
action to the familiar conformal action that we discussed in the
previous section.

By the above discussion it seems that the self adjoint extension is
something beyond our normal understanding of the Bessel process. It
is not difficult to see that it is related to the different possible
boundary conditions for the Bessel process at the origin. Different
self adjoint extensions are related to different boundary conditions
or to different measuring of the paths of the Bessel process. To
make it more rigorous, by Bessel process we mean some kind of
generalized Bessel process which is related to the diffusion of
Brownian motion in the space without the origin. From the two
Brownian motions point of view naively one could argue that since we
have a self adjoint extension just for $\delta<4$ it is just a hint
that two Brownian paths could meet in just $\delta<4$ dimensions. Of
course we are aware that these two statements are not equal, but
they are related because in the case of two Brownian paths one could
think about moving one Brownian motion in the space which remains
after removing from the original space the region filled by another
Brownian motion and the boundary condition on the fixed Brownian
motion is the same as the one that we discussed before. In fact the
interaction of the paths is a point like interaction and as we will
explain soon this interaction plays no role in dimension greater
than three in quantum mechanics, neither for attractive nor for
repulsive interactions.

To understand what could be the meaning of self adjoint extension in
the case of Brownian motion, it is worth to understand the situation
first in the case of quantum mechanics. One way to get a motivation
for the self adjoint extension in the pointed space
$\mathbf{R}^{\delta}$ is by considering the delta function potential
at the origin. Of course this will work just for the integer
$\delta$. The cases $\delta=1,2,3$ are studied with more details for
two reasons, firstly they have more applications and secondly the
$\delta\geq4$ cases are trivial in some senses that we will discuss
soon. The equality of a Hamiltonian with a delta function potential
with a free Hamiltonian on a space with one point deleted plus a
boundary condition were discussed by Jakiew \cite{Jakiew1}. The
equality is in the level of re-normalized delta function which gives
the same scattering data and bound states as produced by the self
adjoint extension. One could re-normalize the delta function with
different functions including sphere delta function, square well and
lattice regularization. For example in the sphere delta function
case, to regularize $V(\mathbf{r})=v\delta(\mathbf{r})$ we can use
\begin{eqnarray}\label{sphere delta}
V(r)=\frac{c_{d}v}{2\pi R^{d-1}}\delta(r-R),
\mbox{with}\hspace{0.5cm}c_{2}=1,\hspace{0.5cm}c_{3}=\frac{1}{2};
\end{eqnarray}
for dimensions $d=2,3$. The corresponding limit is $R\rightarrow0$.
In the case of the square well we have
$V(\mathbf{r})=-D\Theta(R-r)$, where $\Theta$ is the Heaviside
function, $D$ is a coonstant and the limit is the same. Since these
regularization procedures were discussed extensively, we ignore the
details but we explain the strategy, for more details see
\cite{Jakiew1}. Using the above regularized delta potentials firstly
we should solve the problem for the regularized potential by
considering the continuous wave function at $r=R$. This matching
leads to equations between different parameters including $v$, $R$
and energy. Then by defining the re-normalized coupling $\tilde{v}$
as a function of $R$ and $v$, one could derive the energy and wave
function of the system exactly as we found in the previous section.
For example for the three dimensional case we have by definition
$\frac{1}{\tilde{v}}=\frac{1}{v}+\frac{1}{2\pi R}$ and after
matching the wave functions we will have the following equations for
the bound state energy $E_{b}$ and the phase shift of the $s$-wave
scattering sector
\begin{eqnarray}\label{three dimensional bound state and phase shift}
\sqrt{2E_{b}}&=&\frac{2\pi}{\tilde{v}},\\
\tan(\delta_{0})&=&-\frac{\tilde{v}k}{2\pi}.
\end{eqnarray}
Similar results for the two dimensional delta function were given in
\cite{Jakiew1}.

This regularization procedure will not work for dimensions $d\geq4$
because it is not possible to absorb all of the divergent terms in
this case. Although it seems that the re-normalized attractive delta
function in two dimensions has the same physics as the free particle
quantum mechanics in the plane without origin, this matching is not
complete for some reasons. For explanation we need to make some
remarks on the general properties of the delta function potential in
arbitrary dimensions. Firstly the repulsive delta function is
trivial in dimensions higher than one which means that the phase
shift scattering is zero for this case and so the scattering matrix
is equal to one \cite{Friedman}. For the attractive case as we
remarked before it is not possible to define a zero range potential
in more than three dimensions possessing bound states with finite
energy. For simplicity we focus on the three dimensional case and
compare with more detail the delta function potential and the self
adjoint extension counterpart. Let's take the subset of the possible
boundary conditions with the following property for each real
$\eta$,
\begin{eqnarray}\label{boundary condition for d=3}
\psi(0)=\frac{-\eta}{2\pi}\frac{d\psi}{dr}|_{r=0}.
\end{eqnarray}
For positive energy $E=k^{2}/2$ the solution is as follows
\begin{eqnarray}\label{wave function in three dimension}
\psi(r)=\frac{1}{r}(\sin(kr)+\tan(\delta_{0})\cos(kr) ),
\end{eqnarray}
where $\delta_{0}$ is the phase shift corresponding to the $s$-wave.
It is not difficult to see that for the attractive delta function
using equation (\ref{boundary condition for d=3}) and (\ref{wave
function in three dimension}) one could get $\eta=\tilde{v}$. This
matching can not be done for the repulsive case because we already
know that the repulsive delta function is trivial, but the self
adjoint extension with $\eta<0$ still has a phase shift and so it is
not trivial. At the level of the attractive case one may still
interpret the physics of the self adjoint extension for the integer
$\delta$ as the stochastic process in the presence of the
regularized delta function. However ,writing a stochastic equation
for the radial Brownian motion with one removed point is not obvious
and needs more investigation. Of course the definition of a
stochastic process for the generic case with arbitrary $\delta$ or
$\nu$ is more difficult. Although the boundary condition for
$\delta=3$ or $\nu=\frac{1}{2}$, which is equivalent to the half
line free quantum particle, is very simple, it is enlightening. One
could see the boundary condition (\ref{boundary condition for d=3})
from the stochastic process point of view at one dimension as
follows: $\eta=0$ is the Dirichlet boundary condition which is an
absorbing boundary condition. The particle will be absorbed by the
origin after hitting that. This also corresponds to the conventional
free Hamiltonian with scale invariance. From the two particle point
of view it is like absorption of one particle by the other one when
they touch each other. The Dirichlet boundary case is just
reminiscent of the equality of the ~3 dimensional Bessel process
with the Brownian motion on the half line with absorbing boundary
condition.

The other extreme case is the Neumann boundary condition
$\eta\rightarrow-\infty$ which is a reflecting boundary condition at
the origin for the Brownian motion. Other negative values of $\eta$
correspond to a mixing of Dirichlet and Neumann boundary conditions.
This correspondence was discussed in detail in
\cite{clark.menikoff.sharp,gutman1,gutman2} and the Green function
has the following form for arbitrary value of the self adjoint
extension
\begin{eqnarray}\label{green function in three}
G_{\eta}(x,y,t)&=&G_{F}(x-y,t)+G_{F}(x+y,t)+\frac{4\pi}{\eta}\int_{0}^{\infty}dwe^{\frac{2\pi}{\eta}w}G_{F}(x+y+w)\hspace{0.3cm}\eta\leq0;\qquad\\
G_{\eta}(x,y,t)&=&G_{F}(x-y,t)+G_{F}(x+y,t)-\frac{4\pi}{\eta}\int_{0}^{\infty}dwe^{\frac{-2\pi}{\eta}w}G_{F}(x+y-w)\nonumber\\
&+&\frac{4\pi}{\eta}e^{i\frac{2\pi^{2}t}{\eta^{2}}}e^{-\frac{2\pi}{\eta}(x+y)}\hspace{1cm}\eta\geq0;\\
G_{F}(x-y,t)&=&\frac{1}{\sqrt{2\pi it}}e^{i(x-y)^{2}/2t}.
\end{eqnarray}
For the special cases, Dirichlet and Neumann the results are as
follows
\begin{eqnarray}\label{Dirichlet Newman}
G_{\eta=0}(x,y,t)=G_{F}(x-y,t)+G_{F}(x+y,t);\\
G_{\eta\rightarrow-\infty}(x,y,t)=G_{F}(x-y,t)-G_{F}(x+y,t).
\end{eqnarray}
One can use the above equations to get the Green's function of the
Brownian motion in three dimensions when the origin is removed. Of
course one could get the Green's function for the general case by
using an orthogonal eigenvalue expansion for the self adjoint
operator. The Green's function with respect to the solutions of the
Hamiltonian has the following form
\begin{eqnarray}\label{green function}
G_{\eta}(x,y,t)=\int_{0}^{\infty}dke^{-iE_{k}t}\varphi(y)\varphi^{*}(x).
\end{eqnarray}
In the case of $g<\frac{3}{4}$ since we have also a bound state we
need to add new terms coming from the discrete contribution of this
states by adding $\sum_{b}(e^{iE_{b}t}\psi(r)\psi^{*}(r))$ to the
Green's function, where $E_{b}$ is the bound state energy. The
important conclusion is that one could also consider the above
Green's function as sort of a generalized Bessel process in $\delta$
dimensions. In fact one can get the transition density for the above
processes by just Wick rotation. The above generalization of Bessel
process has not been appeared in the mathematical literature. From
the stochastic process point of view one can derive the above
solutions by considering the local time of the process \cite{revuz
yor}. The definition of the local time of the path $\omega$ at the
point $a$ is as follows
\begin{eqnarray}\label{local time}
t_{l}(a):=\frac{1}{2}\lim_{\epsilon\rightarrow0}\int_{0}^{T}\textbf{1}_{x+\epsilon}(B_{s})ds,
\end{eqnarray}
where $\textbf{1}_{x+\epsilon}(B_{s})$ is the indicator for the time
that the process is in the interval $[0,x+\epsilon]$. One could
naively write the above equation as an integral over a delta
function as $t_{l}(\omega,0)=\int_{0}^{T}\delta(x(t))dt$. Using the
above equation the extended transition density for $\delta=3$ is
just the expectation value of $\exp{(\frac{-2\pi}{\eta}t_{l})}$. The
corresponding stochastic process for the free particle in the half
line is called elastic Brownian motion \cite{revuz yor}. To the best
of our knowledge the problem has not been discussed by the
mathematicians for the generic case. For the Bessel process that we
discussed in the second section one just needs to consider $B=0$ in
the equation (\ref{Self adjoint boundary condition1}). In this case
we will recover conformal symmetry for our process again. It will be
also interesting if one could get the results for the Bessel process
with negative dimension by using the quantum mechanics of a free
particle, in particular the equation (\ref{transition density for
negative dimension}).

\section{Generalization of Bessel Process and CQM}\
\setcounter{equation}{0}\

In this section we want to present some possible generalizations of
the Bessel process with the well known quantum mechanical
counterparts such as, Bessel process with constant drift, Bessel
process in the wide sense, Cox, Ingersoll and Ross (CIR) model and
Morse process which are related to a Coulomb potential, conformal
quantum mechanics, radial harmonic oscillator and Morse potential
respectively. Of course not all of the above processes admit
conformal symmetry; they are reasonable perturbations of the
conformal invariant case.
\subsection{Bessel process with constant drift and Coulomb potential}
The definition of the Bessel process with constant drift is given by
the following stochastic process
\begin{eqnarray}\label{drifted bessel}
dR^{\mu}_{t}=(\frac{\delta-1}{2R_{t}}+\mu)dt+dB_{t}.
\end{eqnarray}
From the Feller classification the boundary at the origin has the
same classification as for the standard Bessel process, the infinity
is a natural boundary condition which is attracting if $\mu$ is
positive and non-attracting for a negative drift. The generator of
the process is
\begin{eqnarray}\label{Bessel drift generator}
\mathcal{L}f(x):=\frac{1}{2}f''(x)+(\frac{1}{2x}(\delta-1)+\mu)
f'(x).
\end{eqnarray}
One could map the solution of the equation (\ref{equation in Laplace
space}) to the Schr\"odinger equation of the Coulomb potential. To
do so we need the Liouville transformation
\begin{eqnarray}\label{Liouville transformation}
R_{\lambda}(r)=2^{\frac{1}{4}}(\frac{r}{\sqrt{2}})^{\frac{\delta-1}{2}}e^{\frac{\mu
r}{\sqrt{2}}}G_{\lambda}(\frac{r}{\sqrt{2}}),
\end{eqnarray}
where $r=\sqrt{2}x$. Using the above transformation one could write
the equation (\ref{Bessel drift generator}) as
\begin{eqnarray}\label{Coulomb Schrodinger}
\frac{d^{2}R_{\lambda}}{dr^{2}}+(\frac{c}{r}-\frac{l(l+1)}{r^{2}})R_{\lambda}=-ER_{\lambda},
\end{eqnarray}
where $E=-\lambda+\frac{\mu^{2}}{2}$, $l=\frac{\delta-1}{2}$ and
$c=-\mu\frac{\delta-1}{\sqrt{2}}$. For integer $\delta$ the equation
is related to the energy levels of a Hydrogen atom. The different
aspects of the equation (\ref{Coulomb Schrodinger}) were discussed
in \cite{Vadim Linetsky} including the transition density for the
different values of $\mu$ and $\delta$, here we just mention some
properties. The Bessel process with constant drift admits a discrete
spectrum for all negative values of $\mu$ which is expectable for
those people who are familiar with the Hydrogen atom. For
$0<\delta<1$ the process admits just one trivial bound state but for
$\delta>1$ it has infinitely many bound states. The process also
admits discrete spectrum for the positive values of $\mu$ for
dimensions in the range $0<\delta<1$. In this case $\delta=1$ is the
critical dimension for admitting discrete energies or not.
\subsection{Bessel Process in the Wide sense}

Another very important generalization of the Bessel process is by
considering a Brownian motion in $\delta$ dimensions with a drift of
magnitude $\mu\geq0$ and then taking the radial part as a time
homogeneous diffusion in $[0,\infty)$. It was shown in
\cite{rogerspitman} that the generator of this process is
\begin{eqnarray}\label{generator of wide bessel}
\mathcal{L}f(r):=\frac{1}{2}f''(r)+(\frac{1}{2r}(\delta-1)+h^{-1}_{\nu}(\mu
r)\frac{dh_{\nu}(\mu r)}{dr} )f'(r),
\end{eqnarray}
where
\begin{eqnarray}\label{h}
h_{\nu}(x)=(\frac{2}{x})^{\nu}\Gamma(1+\nu)I_{\nu}(x),\hspace{1cm}\nu=\frac{\delta}{2}-1.
\end{eqnarray}
The transition density of this process is related to the transition
density of the Bessel process $p_{\delta}(t,x,y)$ as follows
\begin{eqnarray}\label{transition density of wide bessel}
p_{\delta,\mu}(t,x,y)=e^{-\mu^{2}t/2}h^{-1}_{\nu}(\mu
x)p_{\delta}(t,x,y)h_{\nu}(\mu y).
\end{eqnarray}
The above process has an interesting time inversion property. It was
shown in \cite{watanabe}, see also \cite{lawi} that if $X_{t}$ with
$t\geq0$ is a diffusion process and $tX_{\frac{1}{t}}$ with $t>0$ is
homogeneous and conservative, no killing in the interior domain,
then both processes are necessarily Bessel processes in the wide
sense up to a time scaled re-parametrization. It is easy to
generalize the above statement to the processes with time inversion
property of degree $\alpha$ by demanding a Markov process with
homogenous $t^{\alpha}X_{\frac{1}{t}}$. They can be obtained by just
considering Bessel processes in the wide sense with appropriate
power. The Hamiltonian corresponding to this process after
transformation $p_{\delta,\mu}(t,x,y)=r^{\nu+1/2}h_{\nu}(\mu
r)Q_{\nu}(\mu r)$ is as follows
\begin{eqnarray}\label{Hamiltonian of wide bessel}
H&=&\frac{1}{2}(p^{2}+V(r))\nonumber\\
V(r)&=&\frac{\nu^{2}-1/4}{r^{2}}+\mu^{2},
\end{eqnarray}
where we used the equality
$\mu^{2}=\frac{3}{4}\mu^{2}+\frac{\mu}{4r}\frac{(\delta
I_{1+\nu}(\mu r)+\mu r I_{2+\nu}(\mu r))}{I_{\nu}(\mu r)}$. Since
the third term does not have any singularity it means that the
general aspects of the above quantum mechanics are the same as the
Bessel process, in particular it will have a self adjoint extension
for the range $\nu<1$. This equality means that basically the above
two distributions have similar spectrum and also it is a very simple
proof for the equation (\ref{transition density of wide bessel}).
One could use all the previous solutions for the self adjoint
Hamiltonian to get the solutions in this case. The consistency of
the inner product is coming from the Doob's $h$-transform. For the
case of Brownian motion with negative drift $-\mu$ one just needs to
replace the modified Bessel function $I_{\nu}(\mu r)$ with
$K_{\nu}(\mu r)$. The same as the Bessel process case here also one
can generalize the Bessel process in the wide sense by considering
all the possible self adjoint solutions of the Hamiltonian
\cite{Hamiltonian of wide bessel} and then using the Doob's
$h$-transform. It will be really interesting to study this
generalized cases with respect to the time inversion.

\subsection{CIR Model and Radial Harmonic Oscillator}

The Cox, Ingersoll and Ross (CIR) model \cite{CoxIngersoll and Ross}
or radial Ornstein- Uhlenbeck process is widely used for interest
rate framework such as some stochastic volatility models
\cite{Heston}. The definition of the CIR family of diffusions is by
the following equation
\begin{eqnarray}\label{CIR}
dN_{t}=(a-bN_{t})dt+c\sqrt{|N_{t}|}dB_{t},
\end{eqnarray}
with $N_{0}\geq0$, $a\geq0$, $c>0$ and $b$ is an arbitrary real
number. It is not difficult to see that $a=\delta$, $b=0$ and $c=2$
is the squared Bessel process. A CIR process can be represented in
terms of the squared Bessel process as follows
\begin{eqnarray}\label{CIR and Bessel}
N_{t}=e^{-bt}Z(\frac{c^{2}}{4b}(e^{bt}-1)),
\end{eqnarray}
where $Z_{t}$ denotes the squared Bessel process of dimension
$\delta=\frac{4a}{c^{2}}$. The above equation can be checked by
using Ito's formula for deterministic time transformation of a
squared Bessel process. Using the above connection one can easily
classify the properties of the origin as a boundary for the process;
we just need to consider $\frac{4a}{c^{2}}$ as the dimension of the
corresponding Bessel process. To make the connection with the radial
harmonic oscillator let's first map the CIR process to the square
root of it $M_{t}=\sqrt{|N_{t}|}$ by using the Ito's formula as
follows
\begin{eqnarray}\label{CIR root and Bessel}
dM_{t}=((\frac{a}{2}-\frac{c^{2}}{8})\frac{1}{M}-\frac{b}{2}M)dt+\frac{c}{2}dB_{t}.
\end{eqnarray}
One can easily map the above process to the radial harmonic
oscillator by the following Hamiltonian
\begin{eqnarray}\label{Harmonic oscillator Hamiltonian}
H=\frac{1}{2}[p^{2}+\frac{\omega^{2}}{4}x^{2}+\frac{k}{x^{2}}+e].
\end{eqnarray}
where $\omega^{2}=b^{2}$,
$k=\frac{1}{2}(a-\frac{c^{2}}{4})(\frac{a}{2}-\frac{c^{2}}{8}-1)$
and $e=-\frac{b}{2}(a-\frac{c^{2}}{4})-b$. The above Hamiltonian is
just the radial part of the harmonic oscillator in $\delta$
dimension. The above calculation shows that by deterministic time
change one can go from scale invariant Bessel process to non-scale
invariant process with stationary solution. This is a hint to
believe that it should be possible to move from a conformal quantum
mechanics to a radial harmonic oscillator by time translation. This
is in fact natural and was done long time ago in \cite{DFF}. The
strategy is as follows: Firstly one could write $SO(2,1)$ generators
in a more familiar form by defining new generators as follows
\begin{eqnarray}\label{SO(2,1)}
S&=&\frac{1}{2}(\frac{1}{a}K-aH),\nonumber\\
R&=&\frac{1}{2}(\frac{1}{a}K+aH),
\end{eqnarray}
where $a$ is a constant and the commutators are as follows
\begin{eqnarray}\label{SO(2,1)commutator}
[D,R]=iS,\hspace{1cm}[S,R]=-iD,\hspace{1cm}[S,D]=-iD.
\end{eqnarray}
The important point is that the operators $D$ and $S$ correspond to
hyperbolic non-compact transformations and $R$ is the generator
corresponding to a compact rotation. Since all of the above
operators are the invariants of the action one could define a
generic operator as
\begin{eqnarray}\label{G}
G=uH+vD+wK,
\end{eqnarray}
as a constant of the motion. Of course it will correspond to compact
rotation in three dimensions if we consider $\Delta=v^{2}-4uw<0$.
This will be important to get a theory with reasonable time
evolution. From now on we will just consider this case. The simplest
example of this kind of operators is $R$ with $\Delta=-1$. The
action of the operator $G$ on a wave function is as follows
\begin{eqnarray}\label{Gsay}
G\mid\Psi(t)>=i(u+vt+wt^{2})\frac{d}{dt}\mid\Psi(t)>
\end{eqnarray}
which by time transformation could be written as
\begin{eqnarray}\label{Gsay}
G\mid\Psi(\tau)>&=&i\frac{d}{d\tau}\mid\Psi(\tau)>\\
\tau=\frac{4w}{\sqrt{-\Delta}}&\{&\arctan(\frac{2wt+v}{-\Delta})-\arctan(\frac{v}{-\Delta})\}.
\end{eqnarray}
In the new parametrization one could think about $G$ as the new time
translation and thus the new Hamiltonian. After some algebra one
could write the following Hamiltonian
\begin{eqnarray}\label{Harmonic oscillator Hamiltonian2}
\tilde{H}=\frac{1}{2}[p^{2}-\frac{\Delta}{4}x^{2}+\frac{g}{x^{2}}],
\end{eqnarray}
as the most general possible Hamiltonian that could be extracted
from the $SO(2,1)$ group with applicable time translation. To
compare with the CIR model one could write $\Delta=-b^{2}$ and $g=k$
which is an indication to believe that the corresponding time
translation in a Bessel process is related to a compact rotation in
a three dimensional space with metric $(-1,-1,1)$. The generators of
rotations in the planes $xy$, $yz$ and $zx$ are $R$, $D$ and $S$
respectively. Different aspects of the self adjoint extension of the
above Hamiltonian were discussed in \cite{pisani} and references
therein and they are quite similar to the case without harmonic
potential.
\subsection{Morse Potential}
In this subsection for the sake of completeness we want to discuss
briefly another related physical model, Schr\"odinger equation with
Morse potential \cite{morse}. This potential is exactly solvable and
has many applications in molecular physics \cite{morse,morse
application}. This system is related to the quantum mechanics of the
radial Harmonic oscillator. One can find the Morse potential by just
making the variable change $u=-2\ln x$ in the Schr\"odinger
equation, $H\psi(x)=E\psi(x)$ with Hamiltonian (\ref{Harmonic
oscillator Hamiltonian}). Then we will have the following
Schr\"odinger equation with the corresponding Hamiltonian
$H^{m}\phi(u)=E^{m}\phi(u)$ where
\begin{eqnarray}\label{Morse Hamiltonian2}
H^{m}=\frac{1}{2}p^{2}+\frac{\omega^{2}}{32}e^{-2u}-\frac{E-e/2}{4}e^{-u},\hspace{1cm}E^{m}=-\frac{1}{4}(\frac{1}{8}+\frac{k}{2}).
\end{eqnarray}
The above equality means that the Morse potential is just the
canonical transformation, i.e. $u=-2\ln x$ and
$p_{u}=-\frac{1}{2}xp_{x}$, of the radial Harmonic oscillator and so
the dynamical symmetry group of the system is still $SO(2,1)$ with
the transformed coordinates \cite{morse conformal}. The other
important issue is that in this case the domain of the quantum
particle is all of the real line and so it is not necessary to worry
about the boundary condition at the origin. It is also easy to
extract the corresponding stochastic process by using the
Hamiltonian as follows
\begin{eqnarray}\label{Stochastic Morse }
dU_{t}=(f e^{-U_{t}}+l)dt+dB_{t},
\end{eqnarray}
where $f=-\frac{\omega}{4}$ and
$l=\frac{1}{2\omega}(E-e/2)+\frac{1}{2}$. It is also possible to
extract the above equation by using equation (\ref{CIR root and
Bessel}) and the Ito's formula plus time re-parametrization. The
connection of the Morse potential to functionals of the Brownian
motion was discussed before in \cite{Ikeda matsumoto}. It is easy to
see by the Feynman-Kac formula that the Kernel of the Morse
Potential is equal to the following expectation in a stochastic
process
\begin{eqnarray}\label{functional brownian Morse }
E[\exp (\lambda ka_{t}-\frac{1}{2}\lambda^{2}A_{t})|B_{t}=y];
\end{eqnarray}
with $\lambda=\frac{\omega}{4}$ and $k=\frac{E-e/2}{\omega}$ and
\begin{eqnarray}\label{functional brownian}
a_{t}=\int_{0}^{t}\exp(B_{s})ds;\hspace{1.5cm}A_{t}=\int_{0}^{t}\exp(2B_{s})ds.
\end{eqnarray}
In \cite{Ikeda matsumoto} the connection of the above process to the
Maass Laplacian \cite{Fay} were also discussed extensively.

\section{Conclusion}\
In this paper we explained many aspects of the connection between
the Bessel process and its possible generalizations on the one side,
and the conformal quantum mechanics and its generalizations on the
other side. The Bessel process as the path integral interpretation
of conformal quantum mechanics has conformal symmetry before
considering the non-Feller boundary conditions of the process which
correspond to the self adjoint extension of the corresponding
quantum mechanics. These boundary conditions could be Feller type if
we consider more generalized stochastic equations. We also discussed
some generalizations of the Bessel process that have interesting
well-known quantum system counterparts. These generalizations are
based on the connection between the Green's function of the quantum
particle and transition density of the Bessel process. Of course
there are also many other systems but we focused on those that have
conformal symmetry as the dynamical symmetry of system. This work
could be extended in many directions including a rigorous study of
the stochastic processes that correspond to the self adjoint
extension of the singular quantum mechanics in the finite domain.
This could be done by using the definition of local time.
Investigating time inversion properties of the generalized Bessel
processes in the wide sense can be useful in classification of time
invertible processes.

Another interesting study could be the study of the process and
quantum mechanics as a system with supersymmetry; in this case we
will have superconformal symmetry as the symmetry of the quantum
mechanics.
\newline
\newline
\newline\textbf{Acknowledgments}

I thank Benjamin Doyon, Shahin Rouhani and Roberto Tateo  for
careful reading of the manuscript and useful comments. I thank also
Sebastian Guttenberg for stimulating discussions and reading the
manuscript. \vspace{1cm}

\section{Appendix A: Self Adjoint Extension of the Hamiltonian
}\label{appdisk}\ \setcounter{equation}{0}\

In this appendix we will summarize Von Neumann-Weyl method of self
adjoint extension for the Hamiltonian operators \cite{pym,simon}.

Consider a Hilbert space $\mathcal{H}$ then an operator
$(A,\mathcal{D}(A))$ defined on $\mathcal{H}$ is said to be densely
defined if the subset $\mathcal{D}(A)$ is dense in $\mathcal{H}$,
i.e., that for any $\psi\in \mathcal{H}$ one can find in
$\mathcal{D}(A)$ a sequence $\phi_{n}$ which converges in norm to
$\psi$, in other words we should have
$\int_{0}^{\infty}|\psi-\phi_{n}|^{2}dx<\epsilon$ for arbitrary
positive $\epsilon$.

The adjoint operator of an operator $H$ with dense domain
$\mathcal{D}(H^{\dag})$ is $H^{\dag}$. The domain
$\mathcal{D}(H^{\dag})$ is the space of functions $\psi$ such that
the linear form $\phi\rightarrow(\psi,H\phi)$ is continuous for the
norm of $\mathcal{H}$ which guaranties the existence of a
$\psi^{\dag}\in\mathcal{H}$ such that
\begin{eqnarray}\label{innerproduct}
(\psi,H\phi)=(\psi^{\dag},\phi).
\end{eqnarray}

Then one may define $H^{\dag}\psi=\psi^{\dag}$. An operator
$(H,\mathcal{D}(H))$ is said to be symmetric or Hermitian if for
$\phi$, $\psi\in\mathcal{D}(H)$ we have $(\phi,H\psi)=(H\phi,\psi)$.
The operator $H$ with the dense domain $\mathcal{D}(H)$ is said to
be self-adjoint if $\mathcal{D}(H^{\dag})=\mathcal{D}(H)$ and
$H^{\dag}=H$.

Definition of the deficiency subspaces $\mathcal{K}_{\pm}$ are by
\begin{eqnarray}\label{deficiency subspaces}
\mathcal{K}_{\pm}=\{\psi\in\mathcal{D}(H^{\dag}),\hspace{0.5cm}H^{\dag}\psi=\pm
i\psi\},
\end{eqnarray}
with dimensions $n_{\pm}$ which are called the deficiency indices of
the operator $H$ and will be denoted by the ordered pair $(n_{+}
,n_{-})$. The following theorem, discovered by Weyl and generalized
by Von Neumann, is the most important result of this appendix.

\textbf{Theorem}: For an operator $H$ with deficiency indices
$(n_{+} ,n_{-})$ there are three possibilities:

~1: If $n_{+} =n_{-}=0$, then $H$ is self-adjoint.

~2: If $n_{+}=n_{-}=n$, then $H$ has infinitely many selfadjoint
extensions, parameterized by a unitary $n\times n$ matrix with
$n^{2}$ real parameters.

~3: If $n_{+}\neq n_{-}$ , then $H$ has no self-adjoint extension.

A relevant example for the above theorem is the Bessel operator
discussed in the paper. The case on the half line was discussed
extensively in section ~2 and we will not discuss it again but the
case in the the finite interval $[0,L]$ is more complicated and
needs to be discussed separately. In this case for $\nu\geq1$ only
one solution is possible and the deficiency indices are $(1,1)$,
while for $0\leq\nu<1$ both solutions are acceptable and the
deficiency indices are $(2,2)$. This is quite natural because for
the Bessel operator in the finite interval we have another boundary
which is like infinite well, so we can expect that the boundary
condition is like the boundary condition of $\delta=3$ case for
$\nu\geq1$. For the case $0\leq\nu<1$ the situation is more subtle
because it is also possible to have some interacting boundary
conditions. In almost all applications the conditions are separated
and so one can look at the non-singular boundary, i.e. $L$, as an
infinite well with one parameter extension as the $\delta=3$ case.
The boundary condition at the origin is just as before. The
interesting point is, since for a differential operator of order $n$
with deficiency indices $(n,n)$, all of its self-adjoint extensions
have discrete spectrums one could argue that for this case all of
the energy levels are discrete. In the other words for the particle
in the finite sphere with origin removed the spectrum of energy is
completely discrete for $0<\delta<4$.

\section{Appendix B:  Boundary Conditions for the Stochastic Equations
}\label{appdisk}\ \setcounter{equation}{0}\

In this appendix we would like to summarize Feller's classification
of possible boundary conditions for the one dimensional stochastic
equation \cite{feller,Ito}. Consider the following equation as our
stochastic equation
\begin{eqnarray}\label{stochastic equation}
dx_{t}=\mu(x_{t})dt+\sigma(x_{t})dB_{t}
\end{eqnarray}
To classify the possible boundary conditions we need to define the
following two functions as the scale function $s(x)$ and speed
measure $m(x)$ as follows
\begin{eqnarray}\label{scale and speed}
s(x):=\exp(-\int^{x}\frac{2\mu(x')}{\sigma^{2}(x')}dx'),\hspace{1cm}m(x):=\frac{2}{\sigma^{2}(x)s(x)}.
\end{eqnarray}
Using the above functions one can define the following four
different functions for diffusion in the interval with endpoints $l$
and $r$
\begin{eqnarray}\label{different functions}
S[x,y]&=&\int_{x}^{y}s(z)dz,\hspace{0.5cm}S(l,y]=\lim_{x\rightarrow l^{+}} S[x,y],\hspace{0.3cm}S[x,r)=\lim_{x\rightarrow l^{-}} S[x,y],\\
M(c,d)&=&\int_{c}^{d}m(x)dx,\hspace{0.5cm}M(l,y]=\lim_{x\rightarrow l^{+}} M[x,y],\hspace{0.3cm}M[x,r)=\lim_{x\rightarrow l^{-}} M[x,y],\\
\sum(l)&=&\int_{l}^{x}S(l,y]m(z)dz,\hspace{0.5cm}\sum(r)=\int_{x}^{r}S[z,r)m(z)dz,\\
N(l)&=&\int_{l}^{x}S[z,x]m(z)dz,\hspace{0.5cm}N(r)=\int_{x}^{r}S[x,z]m(z)dz.
\end{eqnarray}
The boundary classification depends on the behavior of the above
functions and one can put the possible boundary conditions in one of
the following four types for the endpoint $e$:

~1: regular if $\sum(e)$ and $N(e)$ be finite,

~2: exit if $\sum(e)$ be finite and $N(e)$ be infinity,

~3: entrance if $\sum(e)$ be infinite and the $N$ be finite,

~4: natural if $\sum(e)$ and $N(e)$ be infinity.

 For entrance, exit and natural, no boundary conditions are needed
 but  for regular boundary, the conditional probability is not unique
 and dependent on the boundary conditions. An exit boundary can be reached
 from the interior point of the domain with positive probability
 however it is not possible to start the process from the exit
 boundary. An entrance boundary cannot be reached
 from the interior point of the domain but it is possible to start the process from
 the entrance
 boundary. A natural boundary cannot be reached in finite time
 from the interior point of the domain and it is impossible to start the process from
 the natural boundary. A regular boundary is accessible and could be
 reflecting if $m(e)=0$ and sticky if $m(e)>0$. For reflecting
 boundary point the process spends no time in that point but for
 sticky boundary point the process spends a positive amount of time
 at a sticky point. For example $m(e)=\infty$ is called a killing
 boundary condition.

 For the Bessel process the scale function and
 speed measure are given by
\begin{eqnarray}\label{scale and speed Bessel}
s(x)= \left\{
\begin{array}{l l}
  \frac{1}{\nu}x^{\nu}  \quad \mbox{if $\nu\neq0$}\\
  \ln x  \quad \mbox{if $\nu=0$}\\
\end{array} \right .\hspace{1cm}m(x)=\frac{1}{2}x^{-\nu}.
\end{eqnarray}
Different boundary possibilities for the Bessel process were
discussed already in the paper.

\end{document}